\documentclass[twocolumn]{aastex701}

\shorttitle{The effect of $^{16}$O($^{16}$O, n)$^{31}$S rate}
\shortauthors{Xin et al.}
\graphicspath{{./}{figures/}}

\usepackage{courier}
\usepackage{multirow}
\usepackage{threeparttable}
\usepackage{booktabs}
\usepackage{color}
\usepackage[section]{placeins}
\usepackage{graphicx}
\usepackage{subfigure}
\usepackage{parskip}
\usepackage{longtable}
\usepackage{supertabular}
\usepackage{bm}
\usepackage{amsmath}
\usepackage{amssymb}

\begin{document}

\title{Impacts of the $^{16}$O($^{16}$O, n)$^{31}$S reaction rate on the evolution and
nucleosynthesis in Pop III massive stars}

\author[orcid=0000-0003-3646-9356, sname='Xin']{Wenyu Xin}
\affiliation{School of Physics and Astronomy, Beijing Normal University, Beijing 100875, People's Republic of China}
\affiliation{Institute for Frontiers in Astronomy and Astrophysics,
Beijing Normal University, Beijing 102206, People's Republic of China}
\email[show]{xinwenyu16@mails.ucas.ac.cn}  

\author[orcid=0000-0001-9553-0685, sname='Nomoto']{Ken'ichi Nomoto}
\affiliation{Kavli Institute for the Physics and Mathematics of the Universe (WPI), The University of Tokyo Institutes for Advanced Study, \\
The University of Tokyo, Kashiwa, Chiba 277-8583, Japan}
\email[]{nomoto@astron.s.u-tokyo.ac.jp}

\author[0000-0002-3672-2166]{Xianfei Zhang}
\affiliation{School of Physics and Astronomy, Beijing Normal University, Beijing 100875, People's Republic of China}
\affiliation{Institute for Frontiers in Astronomy and Astrophysics,
Beijing Normal University, Beijing 102206, People's Republic of China}
\email[show]{zxf@bnu.edu.cn}

\author[0000-0002-7642-7583]{Shaolan Bi}
\affiliation{School of Physics and Astronomy, Beijing Normal University, Beijing 100875, People's Republic of China}
\affiliation{Institute for Frontiers in Astronomy and Astrophysics,
Beijing Normal University, Beijing 102206, People's Republic of China}
\email[]{bisl@bnu.edu.cn}

\begin{abstract}
We first present a systematic investigation into the effect of the
$^{16}$O($^{16}$O, n)$^{31}$S reaction rate on the evolution and
nucleosynthesis of Population III (Pop III) stars.
We simulate the evolution of a 15 M$_\odot$ Pop III star from the
zero-age main sequence through to core collapse, while varying the
$^{16}$O($^{16}$O, n)$^{31}$S reaction rate by factors of 0.1, 1, and 10.
Our results demonstrate that increasing this reaction rate prompts earlier onset
and extended duration of core oxygen burning at lower temperatures and densities.
A higher reaction rate also increases neutron excess in OSi-rich layers,
thereby promoting the synthesis of neutron-rich isotopes, particularly $^{31}$P and $^{39}$K.
Most notably, the K yield is enhanced by a factor of 6.4.
For a tenfold enhancement of the $^{16}$O($^{16}$O, n)$^{31}$S rate,
the predicted [K/Ca] and [K/Fe] values from presupernova models reach
0.29 and 0.22 dex, respectively—values that are consistent with the most
recent observational data for extremely metal-poor stars.
These findings hold promise as a potential new solution to the problem of
potassium underproduction and offer a valuable theoretical reference 
and motivation for subsequent measurements of oxygen fusion reaction rate.
\end{abstract}



\section{Introduction} \label{sec:intro}

After the Big Bang, the primordial gas was composed primarily of hydrogen and helium,
with minor contributions from light elements such as lithium, beryllium, and boron.
The heavier elements were subsequently synthesized during the evolution and
explosion of the first stars, referred to as Population III (Pop III) or metal-free stars. 
These massive stars concluded their lives as the first supernovae,
thereby injecting large quantities of energy and
newly synthesized metals into the interstellar medium.
This profoundly affected the early galaxies, influencing their dynamics,
thermal properties, and chemical composition.
However, observing these first stars presents a significant challenge,
as they are likely to exist in the most distant reaches of the universe.
Long-lived, low-mass, metal-poor stars serve as crucial observational 
probes for constraining the characteristics of the first stars,
as they preserve the nucleosynthetic imprints of the first supernovae
in their surface abundances \citep{2005ARA&A..43..531B, 2013ARA&A..51..457N,
2022ApJ...931..147L, 2023Natur.618..712X}.

In the early universe, the primordial gas was characterized by
an extremely low metal content, and the neutron excess in Pop III stars
remained nearly zero until the completion of helium burning.
However, the $^{12}$C($^{12}$C, n)$^{23}$Mg reaction plays a crucial role, 
as the frequent decay of $^{23}$Mg to $^{23}$Na enhances the neutron excess
during the carbon-burning phase \citep{Woosley_2002, 2015PhRvL.114y1102B}.
Subsequently, the $^{16}$O($^{16}$O, n)$^{31}$S reaction,
along with various weak interaction processes during the oxygen-burning phase,
can further increase the neutron excess ($\eta$). Given that oxygen
is more abundant than carbon in the carbon-oxygen (CO) core,
a substantial number of protons are converted into neutrons throughout
both the central and off-central regions.
The neutron excess is not only important for nucleosynthesis
during the advanced stellar burning stages \citep{Woosley_2002, 2002ApJ...565..385U},
but also affects the electron mole fraction ($Y_{\rm e}$),
within the star, where $\eta = 1-2\,Y_{\rm e}$.
The value of $Y_{\rm e}$ decreases significantly near the bottom of O burning,
which subsequently impacts the iron (Fe) core mass and final explosion
\citep{Woosley_2002, 2023ApJ...952..155Z, 2025arXiv250211012X}.
Additionally, the $^{16}$O($^{16}$O, n)$^{31}$S reaction may serve
as a neutron source for the s-process in massive stars
\citep{2025NuScT..36...43W, 2025NuScT..36..168Z, 2023PhRvC.107d5809Z}.
The total reaction rate for $^{12}$C+$^{12}$C and its branching rate 
have been measured and constrained by \citet{2015PhRvL.114y1102B} and
China Jinping Underground Laboratory (CJPL; \citealt{2019NuScT..30..126T,
2020PhLB..80135170Z, 2022ChPhC..46j4001W, 2022EPJWC.26001002T, 2024NuScT..35..208N, 
2025PhLB..86239341N, 2025NuScT..36..143W, 2024NuScT..35..217L}),
and the related impact on the stellar evolution and nucleosynthesis has
also been investigated by \citet{2021ApJ...916...79C, 
2023ChPhC..47c4107X, 2025A&A...702A..86D}.
However, for the total $^{16}$O+$^{16}$O and the branching
reaction rates, particularly for $^{16}$O($^{16}$O, n)$^{31}$S,
there have not yet been many experimental measurements and sensitivity investigations.

Previous studies from \citet{Farmer_2020} and \citet{2025A&A...702A..86D}
investigated the effects of the total $^{16}$O+$^{16}$O reaction rate
on the evolution of pulsational pair-instability supernovae (PPISNe)
and the progenitors of the core-collapse supernovae (CCSNe), respectively.
However, these studies did not examine the effects of
individual branching reactions.
\citet{2018ApJS..234...19F} reported that varying the
$^{16}$O($^{16}$O, n)$^{31}$S reaction rate
by a factor of 10 may slightly change the central $Y_{\rm e}$ by approximately 4\%.
However, their investigation was limited to solar and subsolar metallicity stars.
In Pop III stars, the initial neutron excess is exactly zero.
Even a modest change in the neutron excess can exert a pronounced
influence on stellar structural evolution and nucleosynthetic yields.

To illustrate the importance of the $^{16}$O($^{16}$O, n)$^{31}$S reaction rate
and motivate the subsequent experimental measurement,
we systematically examine the effects of
the $^{16}$O($^{16}$O, n)$^{31}$S reaction rate on the evolution and
nucleosynthesis of Pop III stars for the first time.
The paper is organized as follows:
Section \ref{sec:model} describes the adopted stellar models and details
the input assumptions, establishing the basis for our study.
Section \ref{sec:result} presents the simulation results in detail.
Section \ref{sec:conclusion} provides a discussion of the implications of
our findings and summarizes the main conclusions of this work.

\section{Methods and Models}  \label{sec:model}

All calculations in this work were performed using the
Modules for Experiments in Stellar Astrophysics (MESA, version 24.08.1; 
\citet{2011ApJS..192....3P, 2015ApJS..220...15P, 2019ApJS..243...10P, 2023ApJS..265...15J})
to track nuclear burning processes and structural evolution in stars
with $M_{\rm ZAMS}$ = 15 M$_\odot$ from the zero-age main sequence (ZAMS)
until the Fe core collapse, defined as the point where 
infall speed of the Fe core reaches 1000 km s$^{-1}$.
Since our focus is solely on the effects of the $^{16}$O($^{16}$O, n)$^{31}$S
reaction rate, the calculations are terminated prior to the supernova explosion.

\subsection{Presupernova Evolution} \label{sec:pre-ccsn}

The initial chemical composition of Pop III stars is assumed to
reflect primordial Big Bang nucleosynthesis, as adopted from \citet{2007ARNPS..57..463S}.
The mass fractions for $^{1}$H, $^{2}$H, $^{3}$He, $^{4}$He, and $^{7}$Li are 0.7516, 
4.01$\times 10^{-5}$, 2.39$\times 10^{-5}$, 0.2483, and 2.26$\times 10^{-9}$.
Given that Pop III stars are metal-free, stellar wind mass loss
is considered negligible in our calculations.

Convective boundaries in our models are determined using
the Ledoux criterion, and semiconvective mixing is implemented with
an efficiency parameter $\alpha_{\rm sc} =$ 0.01.
Following the simulation of \citet{2011A&A...530A.115B},
step overshooting is applied during core hydrogen burning
with $\alpha_{\rm ov} = 0.335$. After core hydrogen burning,
exponential overshooting is implemented at the top of
core helium burning with $f_{\rm ov} = 0.01$,
and a small degree of overshooting ($f_{\rm ov} = 0.005$) is used at 
the tops of all other convective cores to suppress numerical artifacts.
No overshooting is applied at the base of the shell helium burning.
Because the structure is expected to evolve faster than convection
can reach a steady state in the late burning stages,
we employ the time-dependent convection (TDC, \citet{2008AcA....58..193S}),
which is now included in the MESA code and
described in detail by \citet{2023ApJS..265...15J}.

\begin{table}[htb]
\centering
\caption{Nuclides included in the nuclear reaction network.}
\label{tab:network}
\begin{tabular}{cccccc}
\toprule
Element & $A_{\rm min}$ & $A_{\rm max}$ & Element & $A_{\rm min}$ & $A_{\rm max}$ \\
\midrule
n       & 1    & 1   &   S       & 31   & 35  \\  
H       & 1    & 2   &   Cl      & 35   & 38  \\  
He      & 3    & 4   &   Ar      & 35   & 40  \\  
Li      & 7    & 7   &   K       & 39   & 44  \\  
Be      & 7    & 10  &   Ca      & 39   & 46  \\  
B       & 8    & 11  &   Sc      & 43   & 48  \\  
C       & 12   & 13  &   Ti      & 43   & 51  \\  
N       & 13   & 15  &   V       & 47   & 53  \\  
O       & 14   & 18  &   Cr      & 47   & 57  \\  
F       & 17   & 19  &   Mn      & 51   & 57  \\  
Ne      & 18   & 22  &   Fe      & 51   & 61  \\    
Na      & 21   & 24  &   Co      & 55   & 63  \\    
Mg      & 23   & 26  &   Ni      & 55   & 64  \\    
Al      & 25   & 28  &   Cu      & 59   & 64  \\    
Si      & 27   & 31  &   Zn      & 60   & 64  \\    
P       & 30   & 33  &           &      &     \\  
\bottomrule
\end{tabular}
\end{table}

To calculate nucleosynthesis, we adopt a nuclear network consisting
of 161 isotopes (\texttt{mesa\_161.net}) to follow
the hydrostatic stellar evolution. 
The isotopes included in this network are outlined in Table~\ref{tab:network}.
Following our previous studies \citep{2025arXiv250413766X, 2025arXiv250211012X},
The nuclear reaction rates are taken from the latest version of JINA reaclib
compilation \citep{2010ApJS..189..240C} or the NACRE \citep{1999NuPhA.656....3A}
database when JINA REACLIB data were not available.

\subsection{Reaction Rates in O Burning} \label{sec:oo-rate}

The O burning ignites when the temperature reaches 1.9 GK
\citep{Woosley_2002} and proceeds explosively
over the temperature range of 3 - 4 GK. In the O fusion reaction,
the compound states of $^{32}$S are formed and decay via four channels,

\begin{equation} \label{equ:1616}
\begin{aligned}
^{16}\mathrm{O} + {}^{16}\mathrm{O} \;\rightarrow\; {}^{32}\mathrm{S}^{*} \;\rightarrow\; &^{31}\mathrm{S} + n  +1.45\ \text{MeV} \\
&^{31}\mathrm{P} + p +7.68\ \text{MeV} \\
&^{30}\mathrm{P} + d -2.41\ \text{MeV} \\
&^{28}\mathrm{Si} + \alpha +9.59\ \text{MeV}
\end{aligned}
\end{equation}

\citet{CAUGHLAN1988283} provides the standard and most widely adopted
$^{16}$O+$^{16}$O reaction rate used in stellar evolution calculations.
Owing to the experimental challenges posed by oxygen targets and
the existence of multiple decay channels,
only a limited number of measurements were conducted during the 1980s
\citep{1979PhRvC..20.1305K, 1980ZPhyA.297..161H, 1984NuPhA.422..373W}.
More recent measurements by \citet{2015JPhG...42f5102D}
have yielded only sparse data.
Theoretical predictions of the astrophysical S-factor for
the $^{16}$O+$^{16}$O reaction have been presented by \citet{2020PhRvC.102d4603R}.

\begin{figure}[htbp]
\centering
\begin{minipage}[c]{0.48\textwidth}
\includegraphics [width=80mm]{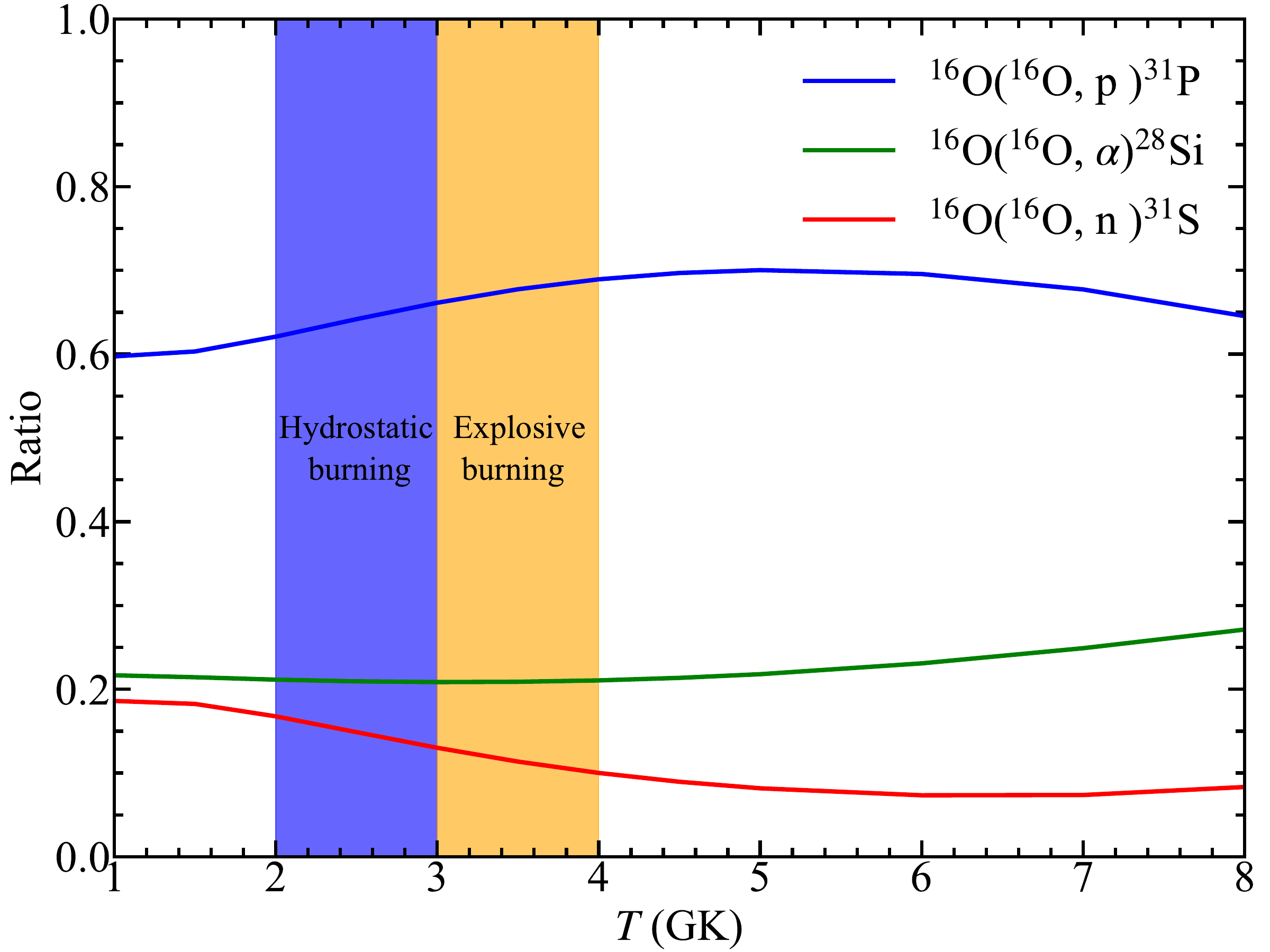}
\end{minipage}%
\caption{The branching ratios of the three branching reactions
as a function of temperature in JINA REACLIB.
The blue and orange regions represent the temperature range of
hydrostatic burning and explosive burning.}
\label{fig:r1616}
\end{figure}

Figure \ref{fig:r1616} presents the branching ratios for neutron,
proton, and $\alpha$ channels as adopted in JINA REACLIB,
noting that the deuteron channel is not included.
The branching ratios during hydrostatic burning
for the neutron, proton, and $\alpha$ channels are 15\%, 65\%,
and 20\%, respectively. At higher temperatures,
the deuteron channel becomes accessible,
with branching ratios for neutron, proton, deuteron,
and $\alpha$ of 5\%, 56\%, 5\%, and 34\%, respectively 
\citep{CAUGHLAN1988283}.

Given the lack of quantified data uncertainties,
we assess the effect of the $^{16}$O($^{16}$O, n)$^{31}$S reaction rate
by scaling it by factors of $f_{\rm 16O} = 0.1$, 1, and 10,
while holding all other branching rates unchanged.
Accordingly, the total $^{16}$O+$^{16}$O reaction rate during hydrostatic burning
is altered by approximately 0.85, 1.00, and 2.51 times, respectively.

\section{Evolution and Nucleosynthesis of Pop III stars}  \label{sec:result}

\subsection{Effect on the Core O burning}  \label{sec:o-b}

\begin{figure}[htbp]
\centering
\begin{minipage}[c]{0.48\textwidth}
\includegraphics [width=80mm]{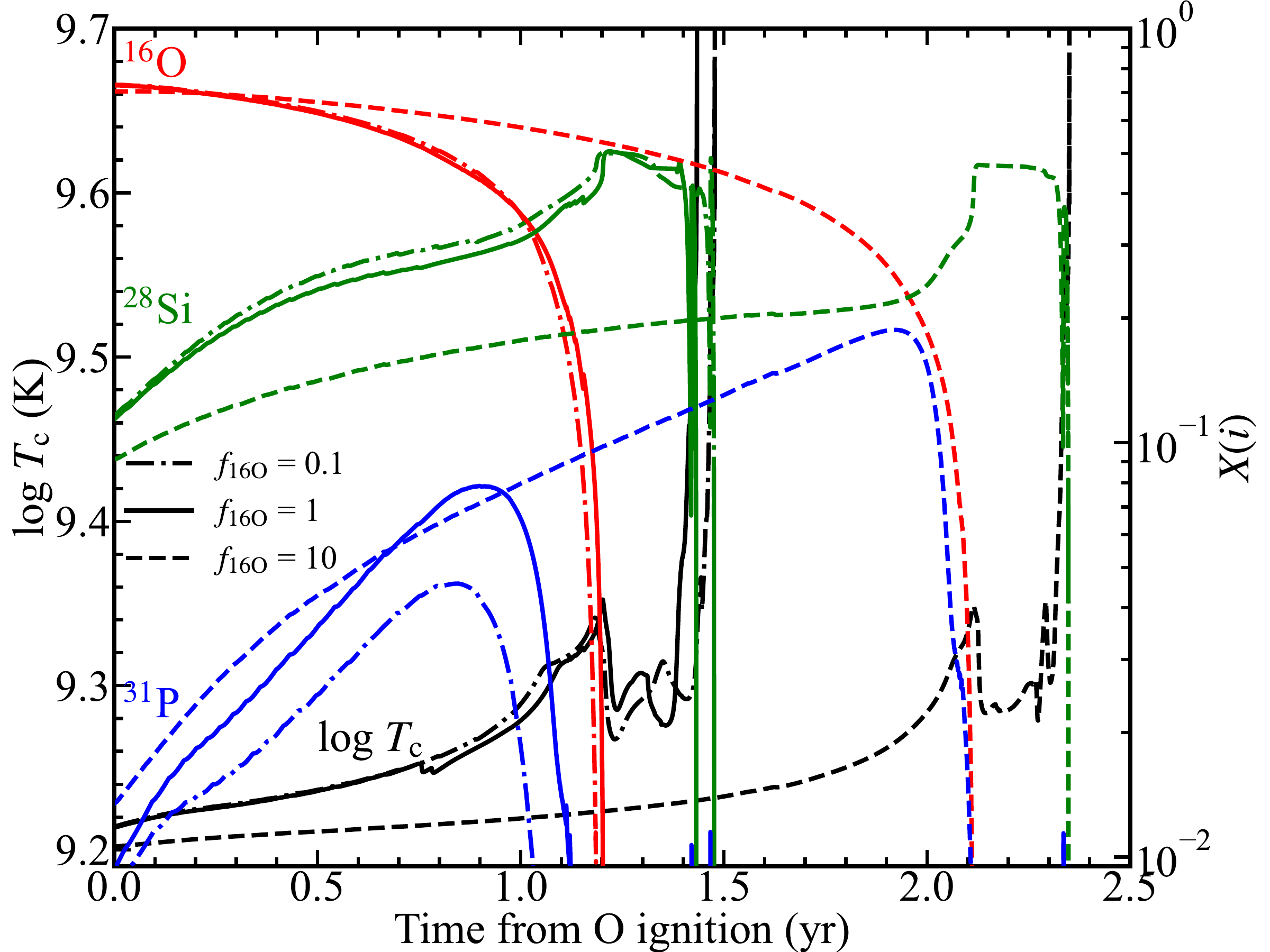}
\end{minipage}%
\caption{The time evolution of temperature, and the mass fractions of $^{16}$O,
$^{28}$Si, and $^{31}$P at the center. The time $t=0$ is defined at O ignition
for each model, where the energy generation rate of $^{16}$O+$^{16}$O reaction
equals the energy loss rate of neutrinos. }
\label{fig:o-b}
\end{figure}

Although the branching ratio of the $^{16}$O($^{16}$O, n)$^{31}$S channel is small, 
increasing this reaction rate nonetheless produces two significant effects.
First, it elevates the overall $^{16}$O+$^{16}$O fusion rate and,
consequently, the nuclear energy generation rate associated with
the $^{16}$O+$^{16}$O reaction, as described by \citet{Woosley_2002},

\begin{equation} \label{equ:eps}
\epsilon_{16,16} \propto X^2(^{16}\rm O) \, \rho \, \lambda_{16,16}
\end{equation}

where $X$($^{16}$O) denotes the mass fraction of $^{16}$O
during oxygen burning and $\lambda_{16,16}$ is the total
$^{16}$O+$^{16}$O fusion reaction rate.
Table \ref{tab:data} demonstrates that models with
$f_{\rm 16O} = 0.1$ and $1$ ignite core oxygen burning at 
nearly identical central temperatures ($\sim$1.62$\times10^9$ K)
and densities ($\sim$3.16$\times10^6$ g cm$^{-3}$),
whereas the model with $f_{\rm 16O} = 10$ ignites at a lower
central temperature ($\sim$1.58$\times10^9$ K) and density
($\sim$3.02$\times10^6$ g cm$^{-3}$).
Figure \ref{fig:o-b} displays the temporal evolution of the central
temperature, as well as the mass fractions of $^{16}$O, $^{28}$Si,
and $^{31}$P. At the onset of core oxygen burning,
the central mass fraction of $^{16}$O decreases from
0.731 to 0.729 and 0.702 as the total $^{16}$O+$^{16}$O reaction rate increases.

\begin{table*}[htb]
\tiny
\centering
\caption{Main data. $T_{\rm O-ign}$ and $\tau_{\rm O-burn}$ represent
the central temperature at O ignition and the lifetime of core O burning.
$X(\rm O)_{\rm max}$ represents the maximum mass fraction of $^{16}$O in the center.
$M(\rm He)$, $M(\rm CO)$, $M(\rm O)$, $M(\rm Si)$, and $M(\rm Fe )$
represent the He core mass, CO core mass, O core mass, Si core mass,
and Fe core mass at $t=t_{\rm final}$, respectively. 
The boundaries of these cores are defined where $X$($^{4}$He), $X$($^{12}$C),
$X$($^{16}$O), and $X$($^{28}$Si) decrease down to 10$^{-4}$, respectively.
$\xi_{2.5}$, log $(V/U_{\rm max})$, and $M(V/U_{\rm max})$
show the parameters for explodability at $t=t_{\rm final}$ and
have been defined by Equations. $Y_{\rm e}$ represent the
central electron mole number at $t=t_{\rm final}$.}
\label{tab:data}
\begin{tabular}{ccccccccccccccccc}
\toprule
$f_{\rm 16O}$ & $T_{\rm O-ign}$ & $\rho_{\rm O-ign}$ & $\tau_{\rm O-burn}$ & $X(\rm O)_{\rm max}$ & $M(\rm He)$ & $M(\rm CO)$ & $M(\rm O)$  & $M(\rm Si)$  & $M(\rm Fe )$&$\xi_{2.5}$ & log $(V/U_{\rm max})$ &  $M(V/U_{\rm max})$ &$Y_{\rm e}$ \\
              &   (K)     &   (g cm$^{-3}$)   &  (yr)   &            &  (M$_\odot$)& (M$_\odot$) & (M$_\odot$) & (M$_\odot$)  &  (M$_\odot$) &            &            &     (M$_\odot$)    &         &            \\
\midrule
0.1   & 1.62$\times10^9$ & 3.16$\times10^6$  & 1.19 & 0.731 &  4.30 & 3.64 & 1.81 & 1.51  & 1.00 & 0.159   & 1.52  &  1.56   &0.434\\  
1.0   & 1.62$\times10^9$ & 3.16$\times10^6$  & 1.20 & 0.729 &  4.30 & 3.65 & 1.80 & 1.46  & 1.02 & 0.140   & 1.54  &  1.52   &0.433\\  
10.0  & 1.58$\times10^9$ & 3.02$\times10^6$  & 2.10 & 0.702 &  4.30 & 3.64 & 1.62 & 1.43  & 1.04 & 0.144   & 1.58  &  1.52   &0.434\\  
\bottomrule
\end{tabular}
\end{table*}

According to Equation \ref{equ:eps}, $\epsilon_{16,16}$ is enhanced by factors 
of 0.85, 1.00, and 2.22 for $f_{\rm 16O}$ = 0.1, 1, and 10, respectively.
Relative to the default value ($f_{\rm 16O} = 1$),
reducing the $^{16}$O($^{16}$O, n)$^{31}$S reaction rate by
a factor of 10 decreases $\lambda_{16,16}$ and
$\epsilon_{16,16}$ by only about 15\%,
whereas increasing the rate by a factor of 10 results in more
than a twofold increase. The evolution of the central temperature and of
$X$($^{16}$O) and $X$($^{28}$Si) remains largely unchanged
when this rate is reduced by a factor of 10,
and the core oxygen burning lifetime is essentially unaffected.
In contrast, a tenfold increase in this rate significantly increases
$\lambda_{16,16}$ and $\epsilon_{16,16}$,
causing oxygen burning to occur at lower temperatures and densities.
Oxygen is consumed more slowly,
extending the core oxygen burning lifetime to 2.10 years.
We also find that the maximum $X$($^{31}$P),
which is the decay product of $^{16}$O($^{16}$O, n)$^{31}$S,
is reduced by 40\% or enhanced by more than a factor of 2 for
$f_{\rm 16O} = 0.1$ and 10, respectively.

In conclusion, increasing the $^{16}$O($^{16}$O, n)$^{31}$S reaction rate
by a factor of 10 leads to an earlier onset of core oxygen burning.
Under these conditions, the burning proceeds at lower temperatures and densities,
resulting in an extended core oxygen burning lifetime.
Furthermore, the peak value of $X$($^{31}$P) is enhanced by more
than a factor of two.

\subsection{Effect on shell O burning and the final structure}  \label{sec:exp}

\begin{figure}[htbp]
\centering
\begin{minipage}[c]{0.48\textwidth}
\includegraphics [width=80mm]{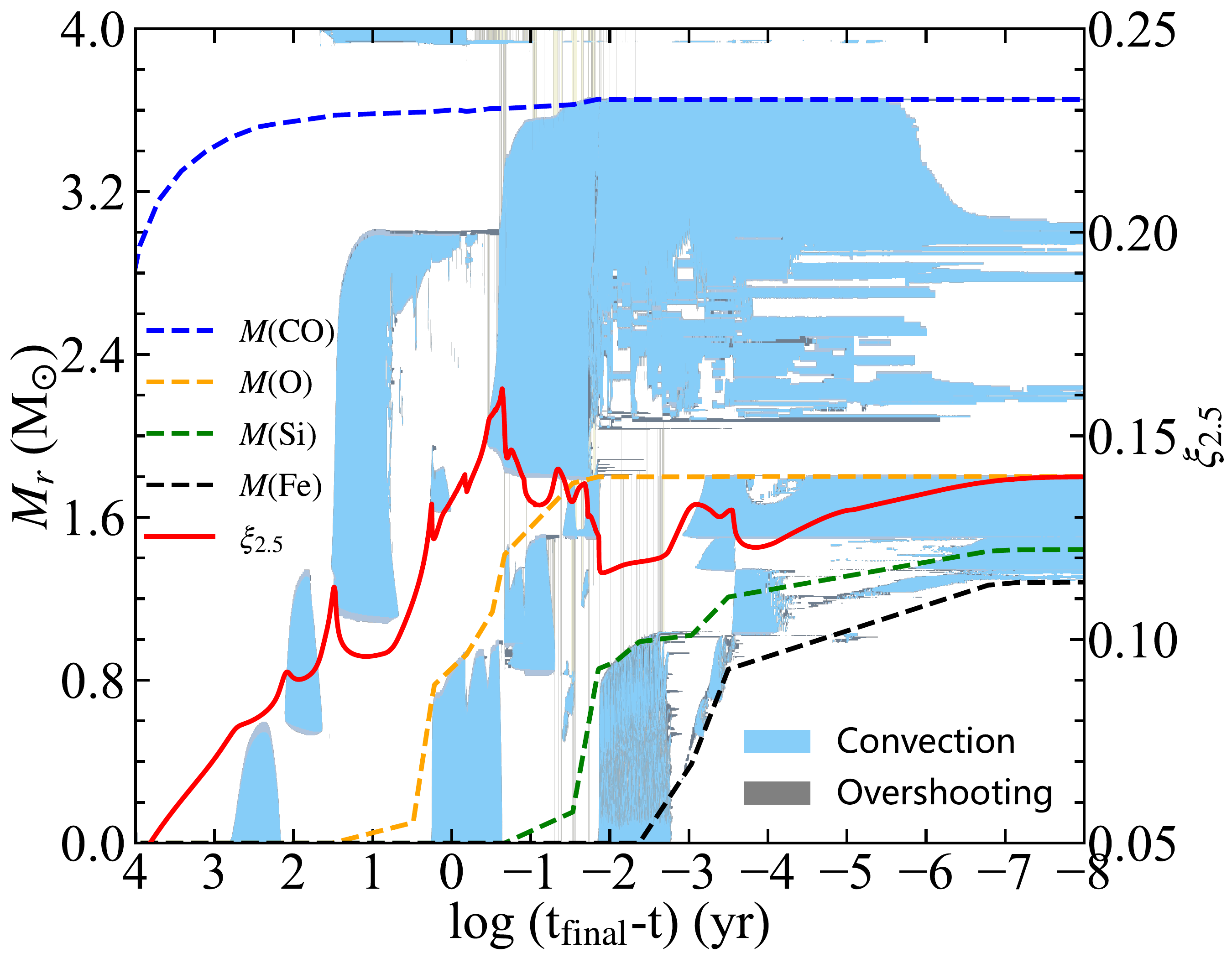}
\end{minipage}%
\caption{The Kippenhahn diagram of the star with $M {\rm (ZAMS)}$ = 15 M$_{\odot}$.
The inner part of $M_r = 0 - 4$ M$_{\odot}$ is shown.
The blue and grey regions represent the convection and overshooting.
The blue, orange, green, and black dashed lines show the variation of CO core
mass, O core mass, Si core mass, and Fe core mass. The boundaries of these cores are
defined where $X$($^{4}$He), $X$($^{12}$C), $X$($^{16}$O), and $X$($^{28}$Si)
decrease down to 10$^{-4}$.}
\label{fig:kipp_M15}
\end{figure}

Figure \ref{fig:kipp_M15} presents the Kippenhahn diagram,
depicting the evolution of core masses from the end of core helium 
burning to the onset of Fe core collapse for $M {\rm (ZAMS)}$ = 15
M$_{\odot}$ and the default reaction rate ($f_{\rm 16O} = 1$).
Here, $t_{\rm final}$ is defined as the moment when the infall speed 
of the Fe core reaches 1000 km s$^{-1}$.
The compactness parameter is defined at $M_r$ = 2.5 M$_\odot$ and
used to evaluate the explodability \citep{2011APS..DNP.CG005O},

\begin{equation} \label{equ:compact}
\xi_{M_r} = \frac{M_r/{\rm M}_{\odot}}{R(M_r)/1000\,{\rm km}}
\end{equation}
where $M_r$ is the enclosed mass at the radius of $R(M_r)$.

After core helium burning, $\xi_{2.5}$ increases as the core contracts.
The residual carbon mass fraction in the core, $X$($^{12}$C) = 0.285,
is sufficient to ignite central carbon burning.
During the central carbon burning phase, the growth of $\xi_{2.5}$ slows down,
as the energy generated by the convective core retards further contraction.
After core carbon burning, the core contracts rapidly,
triggering vigorous off-center carbon burning. As contraction proceeds,
carbon burning is repeatedly ignited in the shell layers,
becoming progressively more energetic. The convective shell extends outward,
transporting fresh C to its base to sustain burning.
Conversely, the increasingly intense shell carbon burning releases
sufficient energy to inhibit further contraction or even drive expansion.
Subsequent shell oxygen and silicon burning produce the 'knees'
in the evolution track of $\xi_{2.5}$.
This evolutionary behavior is discussed in detail by
\citet{2020ApJ...890...43C} and \citet{2025arXiv250211012X}.

\begin{figure}[htbp]
\centering
\begin{minipage}[c]{0.48\textwidth}
\includegraphics [width=80mm]{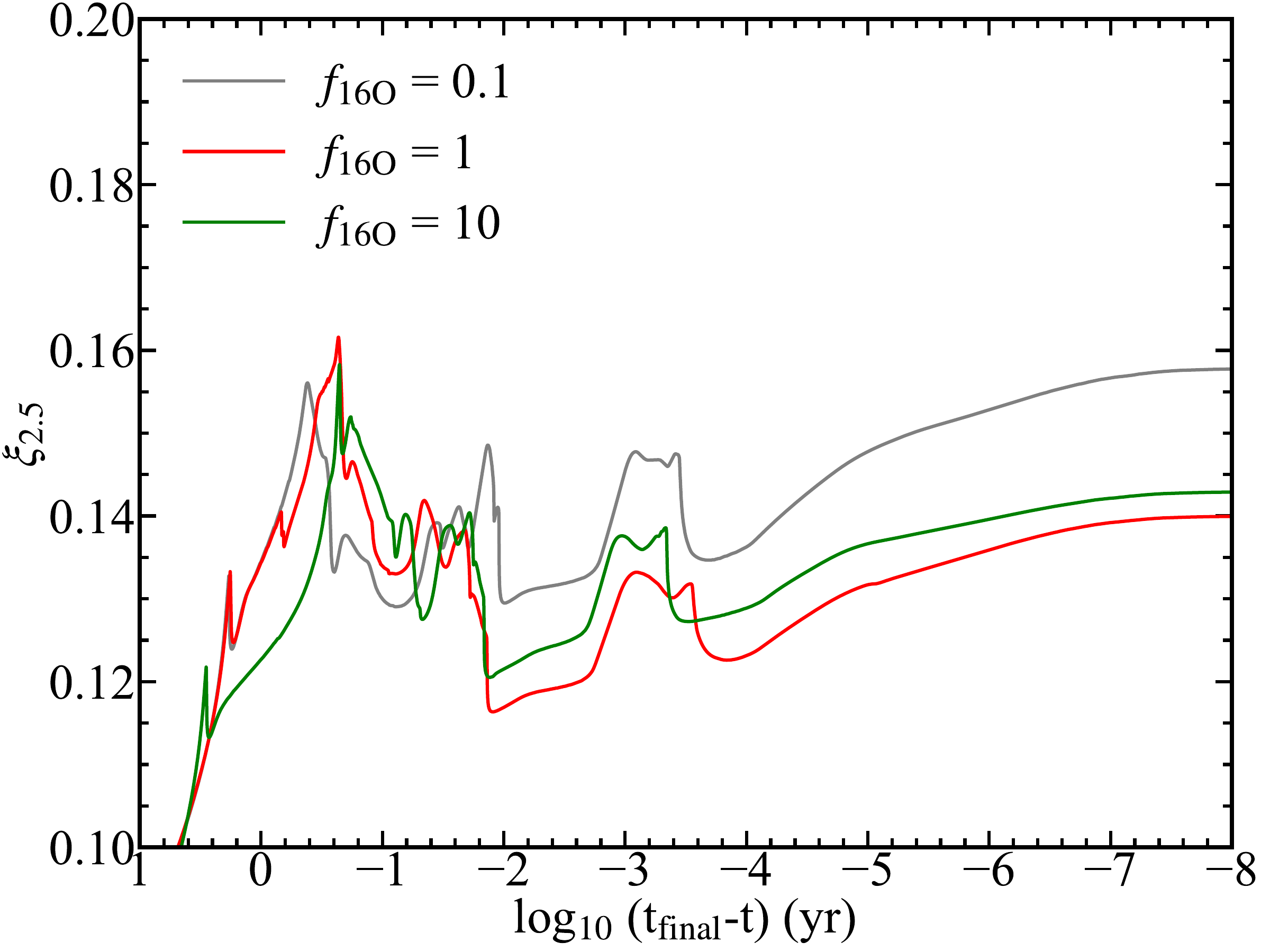}
\end{minipage}%
\caption{The time evolution of compactness parameter, $\xi_{2.5}$, for
$f_{\rm 16O}$ = 0.1, 1, and 10.}
\label{fig:xi}
\end{figure}

Figure \ref{fig:xi} presents the temporal evolution of $\xi_{2.5}$ for
various $f_{\rm 16O}$ values. Based on the discussion of
Figure \ref{fig:kipp_M15}, the sharp peaks appearing between
log$_{10}(t_{\rm final}) - t$ = 1 and 0 are produced by core oxygen burning.
Accordingly, we find that an earlier onset of core oxygen burning
(higher $f_{\rm 16O}$) suppresses the increase in $\xi_{2.5}$ and results
in a less compact oxygen core. Nevertheless,
models with $f_{\rm 16O} = 10$ predict a denser core than those with
$f_{\rm 16O} = 1$ at the onset of core collapse.
The oscillatory behavior of $\xi_{2.5}$ during advanced burning stages
is governed by subsequent core and shell burning episodes,
particularly shell carbon burning.
Thus, the final value of $\xi_{2.5}$ reflects the cumulative effects
of multiple shell and core burning phases,
making it challenging to isolate the specific impact of shell
oxygen burning on the final structure based solely on $\xi_{2.5}$.

\begin{figure}[htbp]
\centering
\begin{minipage}[c]{0.48\textwidth}
\includegraphics [width=80mm]{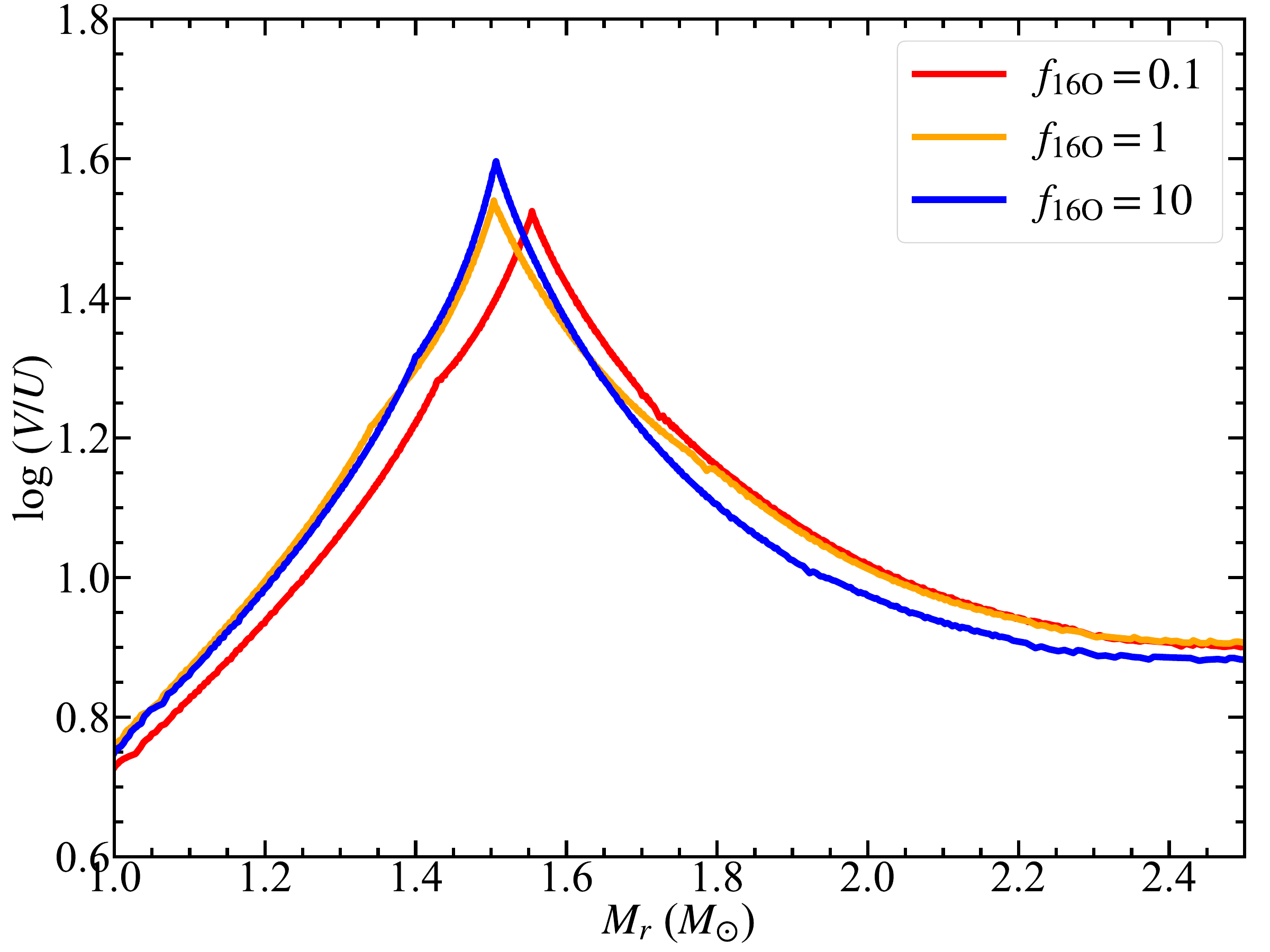}
\end{minipage}%
\caption{The mass distribution of log ($V/U$) at $t=t_{\rm final}$.
The mass coordinates of the peaks indicate $M(V/U_{\rm max})$.
Both the peak values and their mass coordinates are
listed in Table \ref{tab:data}.}
\label{fig:vu}
\end{figure}

In Table \ref{tab:data}, we also evaluate two non-dimensional parameters of
log ($V/U_{\rm max}$) $-$ $M(V/U_{\rm max})$. $V/U$ is defined as

\begin{equation} \label{equ:uv}
\frac{V}{U} = - \frac{{\rm dln} P}{{\rm dln} M_r} = \frac{G M_r^2}{4\pi r^4P}
\end{equation}                                                             
where $U$ and $V$ are defined and discussed in earlier studies
\citep{1980SSRv...25..155S, 2013sse..book.....K}.
The mass distribution of log ($V/U$) is shown in Figure \ref{fig:vu}.
log ($V/U_{\rm max}$) represents the peak of log ($V/U$) and  $M(V/U_{\rm max})$ 
indicates the mass coordinate of log ($V/U_{\rm max}$).
\citet{2025arXiv250211012X} indicated that log ($V/U_{\rm max}$) represents the
strength of the shell O burning. We find that the log ($V/U_{\rm max}$) increases
with $f_{\rm 16O}$ increasing. This implies that the strength of shell O burning
becomes stronger at the onset of core collapse.
As a result, higher $f_{\rm 16O}$ predicts a less compact OSi core and smaller
mass cut, $M(V/U_{\rm max})$, before the explosion.

\begin{figure}[htbp]
\centering
\begin{minipage}[c]{0.48\textwidth}
\includegraphics [width=80mm]{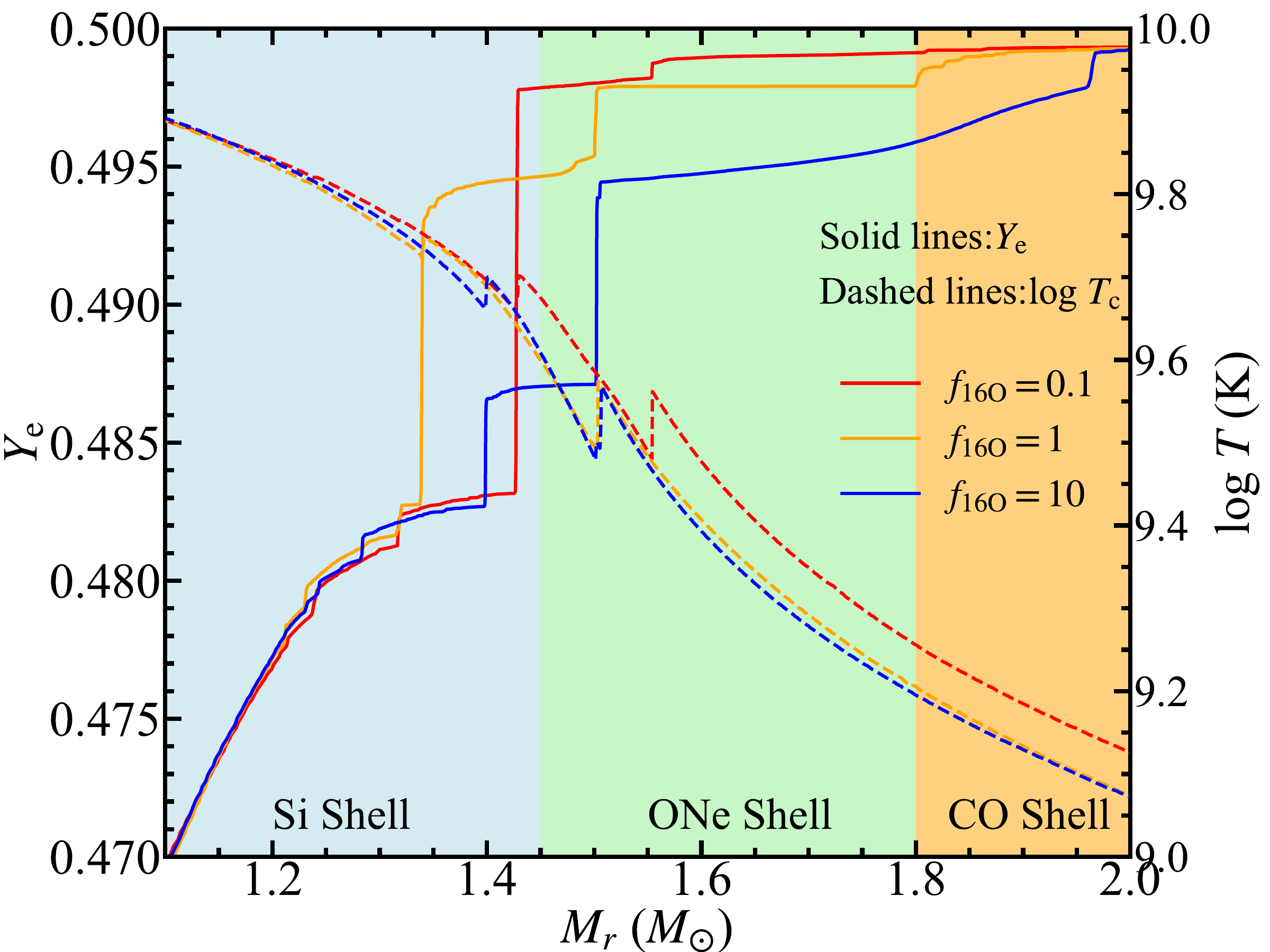}
\end{minipage}%
\caption{The mass distribution of $Y_{\rm e}$ and temperature
in the center for different $f_{\rm 16O}$ at $t=t_{\rm final}$.
The light blue, light green, and orange regions represent the Si,
ONe, and CO layers.}
\label{fig:ye}
\end{figure}

On the other hand, increasing the $^{16}$O($^{16}$O, n)$^{31}$S
reaction rate produces more $^{31}$S,
which subsequently decays to $^{31}$P (see Figure \ref{fig:o-b}).
This decay process alters the electron fraction $Y_{\rm e}$ during
the oxygen-burning phase. Figure \ref{fig:ye} illustrates the mass
distribution of central $Y_{\rm e}$ and temperature for different
$f_{\rm 16O}$ values. In the Fe core and the base of the silicon shell
($M_r \lesssim$ 1.35 M$_\odot$), variations in the 
$^{16}$O($^{16}$O, n)$^{31}$S reaction rate do not affect $Y_{\rm e}$, 
as the temperature in this region is sufficiently high to achieve
complete silicon burning and nuclear statistical equilibrium (NSE),
which are independent of individual reaction rates.
However, this rate significantly impacts $Y_{\rm e}$ from the top of
the silicon-rich layers up to the ONe shell
($1.35 \lesssim M_r \lesssim$ 1.8 M$_\odot$). In these regions,
$Y_{\rm e}$ decreases markedly as $f_{\rm 16O}$ increases.
The $Y_{\rm e}$ discontinuity typically occurs near the base of the
oxygen-burning shell. For $f_{\rm 16O} = 1$,
we observe a $Y_{\rm e}$ jump within the silicon-rich layers,
attributable to the merger of the silicon-burning shell with the
oxygen-burning shell (see Figure \ref{fig:kipp_M15}).
The CO shell ($M_r \gtrsim$ 1.8 M$_\odot$) is largely insensitive
to this reaction rate because temperatures are not sufficiently high
to ignite oxygen burning. Although a decrease in $Y_{\rm e}$ is
observed for $f_{\rm 16O} = 10$, we also note an outward extension of
convective shell oxygen burning in this model prior to core collapse,
which subsequently mixes lower-$Y_{\rm e}$ material outward.
The altered $Y_{\rm e}$ distribution in the ONe and portions of the
silicon-rich layers will also influence the resulting nucleosynthesis.

\subsection{Effect of reaction rate on nucleosynthesis}  \label{sec:nuc}

\begin{figure}[htbp]
\centering
\begin{minipage}[c]{0.48\textwidth}
\includegraphics [width=80mm]{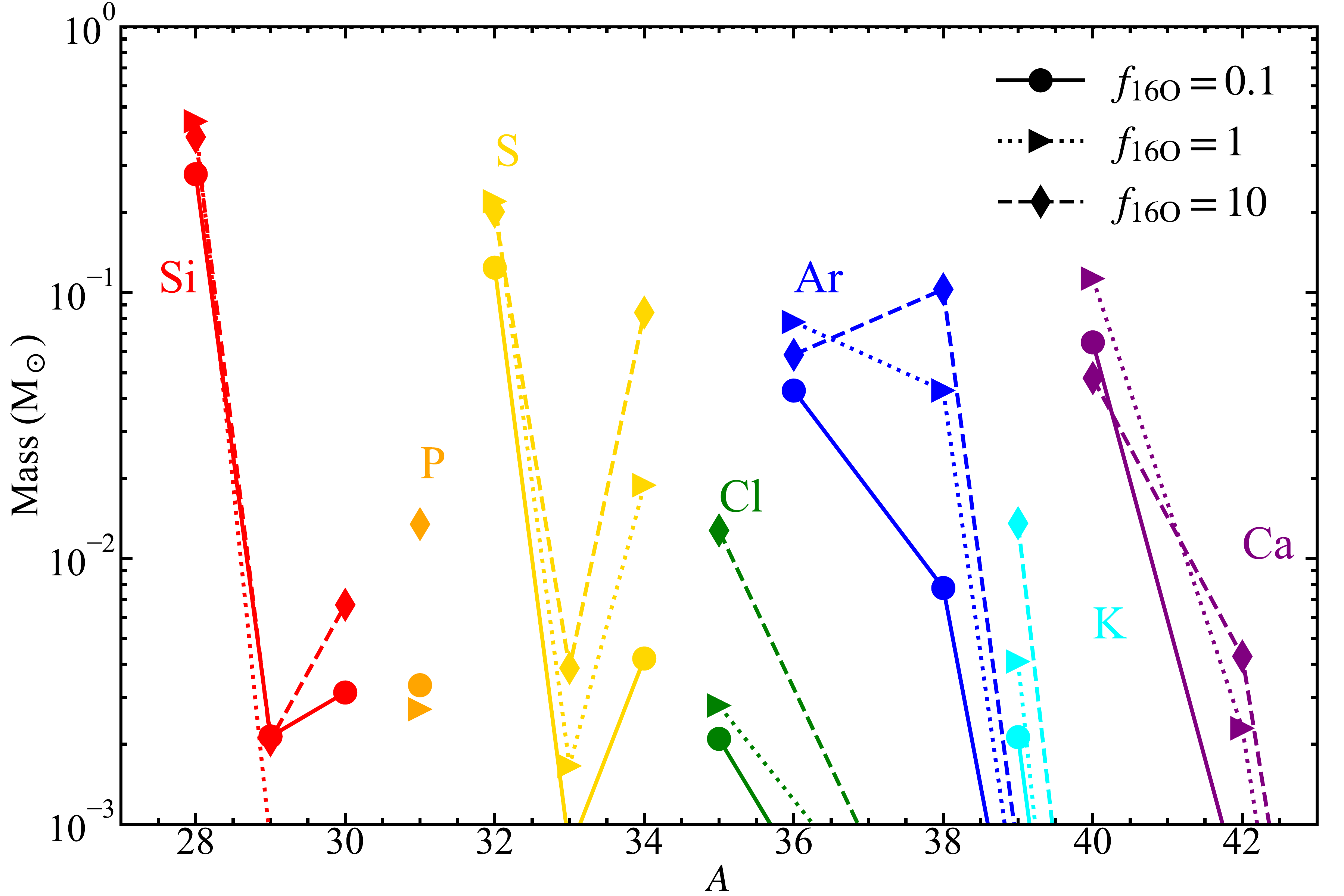}
\end{minipage}%
\caption{The pre-supernova yields of isotopes from Si to Ca
for $f_{\rm 16O}$ = 0.1, 1, and 10 at $t=t_{\rm final}$. }
\label{fig:iso_ratio}
\end{figure}

Approximately 90\% of the products of oxygen burning are $^{28}$Si
and $^{32}$S. The remaining products include $^{33,34}$S, $^{35,37}$Cl,
$^{36,38}$Ar, $^{39,41}$K, and $^{40,42}$Ca.
Figure \ref{fig:iso_ratio} presents the isotope yields from Si to Ca for
$f_{\rm 16O}$ = 0.1, 1, and 10 at $t = t_{\rm final}$.
Elements heavier than Ca are predominantly synthesized during silicon
burning and the NSE process, which are not significantly
affected by the $^{16}$O($^{16}$O, n)$^{31}$S reaction rate.
We assume no fallback during the explosion, and all material
outside the mass cut—evaluated as $M(V/U_{\rm max})$—is ejected.
While explosive burning during the supernova will also modify the
yields of these isotopes, pre-supernova yields are sufficient to
assess the sensitivity to the $^{16}$O($^{16}$O, n)$^{31}$S reaction rate.

Slight variations in the yields of $^{28}$Si, $^{32}$S, $^{36}$Ar, 
and $^{40}$Ca may arise from temperature differences during oxygen
burning and the extent of convective regions.
We find that most neutron-rich isotopes—such as $^{31}$P, $^{33,34}$S,
$^{35,37}$Cl, $^{38}$Ar, $^{39}$K, and $^{42}$Ca—are sensitive to changes
in the $^{16}$O($^{16}$O, n)$^{31}$S reaction rate.
The yields of these isotopes are enhanced by approximately 5 to 10 times
as the reaction rate is varied by two orders of magnitude. In contrast, 
other neutron-rich isotopes are relatively insensitive to this reaction rate, 
since $^{40}$K and $^{46}$Ca are produced during carbon and neon burning, 
and $^{44}$Ca originates almost entirely from the decay of
radioactive $^{44}$Ti formed in the $\alpha$-rich freezeout
\citep{1995ApJS..101..181W}. Note that the yields of
some aforementioned isotopes are below $10^{-3}$ and are
not displayed in Figure \ref{fig:iso_ratio}.

Among these products, our primary focus is on potassium (K). 
K, principally in the form of $^{39}$K,
is synthesized predominantly during the oxygen-burning phase in
massive stars and its yield is further affected by explosive
nucleosynthesis during supernovae. Nevertheless, 
K abundances predicted by Galactic Chemical Evolution (GCE) models
fall short of observations by more than 1 dex at all metallicities.
This discrepancy is chiefly attributed to the underproduction of
K yields in one-dimensional models of massive-star evolution and
supernova explosions \citep{1995ApJS...98..617T, 2013ARA&A..51..457N, 2020ApJ...900..179K}.
Previous studies have shown that potassium abundances can be
elevated to levels consistent with observations under specific circumstances.
For instance, \citet{2018ApJS..237...13L} demonstrated that a model
with an initial mass of 15 M$_\odot$ and a rotational velocity of
300 km s$^{-1}$ predicted enhanced K production.
Furthermore, \citet{2024ApJS..270...28R} noted that,
in cases of carbon–oxygen (C–O) shell mergers,
oxygen-burning products are less likely to be further processed
during explosive nucleosynthesis and are instead ejected,
thereby increasing potassium yields irrespective of metallicity.
In this work, we identify an alternative mechanism for enhancing
the products of oxygen burning, particularly potassium.
As shown in Figure \ref{fig:iso_ratio},
the yield of $^{39}$K increases by approximately 6.4 times,
which we attribute to the higher neutron excess established
in the oxygen-burning shells.

\begin{figure}[htbp]
\centering
\begin{minipage}[c]{0.48\textwidth}
\includegraphics [width=80mm]{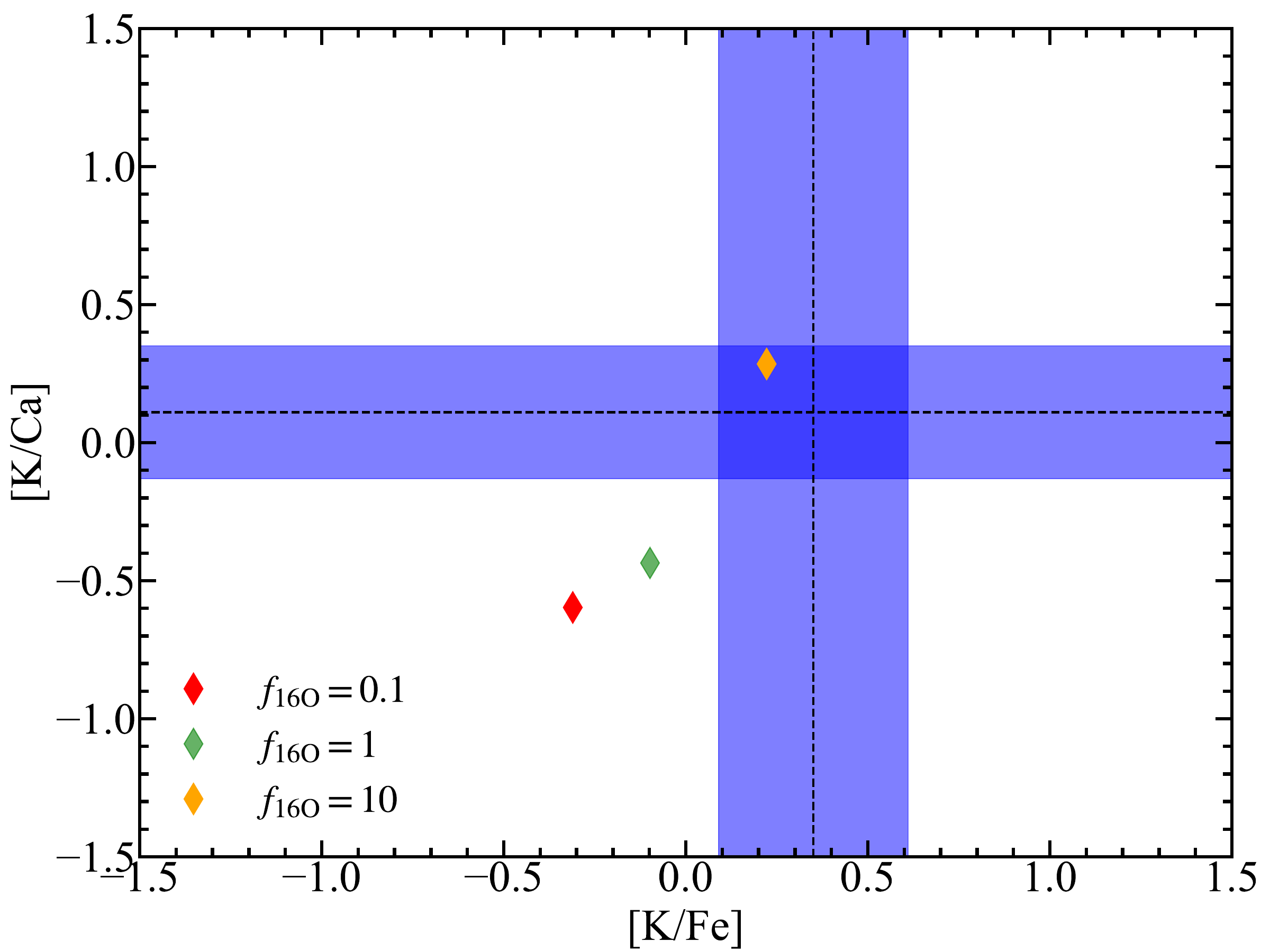}
\end{minipage}%
\caption{The abundance ratios of [K/Ca] against [K/Fe]
for $f_{\rm 16O}$ = 0.1, 1, and 10 before the explosion.
The mean of observed abundances and their 2$\sigma$ reported
in \citet{2025ApJ...992..215I} are shown by the black dashed
lines and the blue regions.}
\label{fig:observe}
\end{figure}

To facilitate comparison with astronomical observations.
We show the abundance ratio of element i relative to element j.
The abundance ratio is defined as:
\begin{equation} \label{equ:x_fe}
[\frac{N_i}{N_j}] \equiv {\rm log}(\frac{N_i}{N_j})- {\rm log}((\frac{N_i}{N_j})_\odot)
\end{equation}
where $N_i$ and $N_j$ represent the number of elements i and j, respectively.
The solar abundance adopted here is from \citet{2009LanB...4B..712L}. 

Figure \ref{fig:observe} displays the abundance ratios [K/Ca] versus [K/Fe].
A sample of extremely metal-poor (EMP) stars with measured K abundances,
based on high-resolution spectra, was reported by \citet{2025ApJ...992..215I}.
The black dashed lines indicate the mean values of these observed abundances,
with [K/Ca] = 0.35 and [K/Fe] = 0.11,
while the blue regions represent their 2$\sigma$ intervals.
Our findings show that the models with $f_{\rm 16O}$ = 0.1 and 1
fall outside the 2$\sigma$ region,
whereas only the model with $f_{\rm 16O}$ = 10 closely matches the observations.
The dominant calcium isotope, $^{40}$Ca,
is largely insensitive to variations in the $^{16}$O($^{16}$O, n)$^{31}$S
reaction rate (see Figure \ref{fig:iso_ratio}).
Although the iron yield, originating from $^{56}$Ni, is sensitive to $Y_{\rm e}$, 
this electron fraction is further modified during silicon burning and
the NSE process, which are not influenced by changes in the
$^{16}$O($^{16}$O, n)$^{31}$S reaction rate.

Consequently, the iron yield is not significantly affected by
variations in this reaction rate.
The predicted [K/Ca] values are -0.60, -0.44, and 0.29,
while [K/Fe] values are -0.31, -0.10, and 0.22 for
$f_{\rm 16O}$ = 0.1, 1, and 10, respectively.
It should be emphasized that these abundance ratios are derived from
presupernova models; explosive nucleosynthesis triggered by the
supernova shock will further modify the yields of K and Ca, and
enhance the Fe yield. However,
the detailed explosion mechanism of core-collapse supernovae,
particularly the role of the neutrino process,
remains uncertain \citep{2008ApJ...672.1043Y, 2024PhRvD.109h3023Z}
and is beyond the scope of this study.

\section{Conclusion}  \label{sec:conclusion}

In this study, we have systematically investigated the impact
of the $^{16}$O($^{16}$O, n)$^{31}$S reaction rate on the late-stage
evolution and nucleosynthesis of massive Pop III stars with
an initial mass of $M(\rm ZAMS)$ = 15 M$_\odot$.
All stellar models are evolved from the zero-age main sequence (ZAMS)
to the onset of Fe core collapse, defined as the point at 
which the infall velocity of the Fe core attains 10$^8$ cm s$^{-1}$.
We assume that all material outside the mass cut ($M_r> M(V/U_{\rm max})$)
is ejected and contributes to the chemical enrichment of the Galaxy.

By varying the reaction rate by factors of 0.1, 1, and 10,
we find that an increased rate results in earlier ignition of
core oxygen burning.
In the case where $f_{\rm 16O}=10$, oxygen burning proceeds at
lower temperatures and densities,
resulting in an extended core oxygen burning lifetime.

Our results also indicate that the $^{16}$O($^{16}$O, n)$^{31}$S
reaction rate significantly affects shell oxygen burning.
An increased reaction rate enhances the strength of shell oxygen burning,
leading to a less compact OSi core and a smaller mass cut before the explosion,
while also increasing the neutron excess in OSi-rich layers and significantly
boosting the production of neutron-rich isotopes, particularly $^{31}$P and $^{39}$K.

Notably, the K yield is amplified by up to a factor of 6.4,
bringing the theoretical [K/Ca] and [K/Fe] abundance ratios into better agreement 
with recent observations of extremely metal-poor stars within 2$\sigma$.
This result provides a new possible solution to the long-standing problem of the underproduction of oxygen burning products, particularly for K.

The abundance ratios of [K/Ca] and [K/Fe] should also be altered by the
explosive nucleosynthesis triggered by the supernova shock.
But the exact explosion mechanism of core-collapse supernovae has
not been well understood.
In future studies, the correction of explosive nucleosynthesis should be
included and related uncertainties should also be investigated.
On the other hand, the more exact $^{16}$O($^{16}$O, n)$^{31}$S reaction
rate is also important for calculating reliable yields of oxygen burning
products.

\begin{acknowledgments}
This work was Supported by the National Natural Science Foundation of China (No.12473028, 12073006, 12090040
and 12090042). W. Y. X. is supported by the Cultivation Project for LAMOST Scientific Payoff
and Research Achievement.
K. N. is supported by the World Premier International Research Center Initiative
(WPI), MEXT, Japan, and the Japan Society for the Promotion of Science
JSPS KAKENHI Grant Numbers JP20K04024, JP21H044pp, JP23K03452, and JP25K01046.
\end{acknowledgments}

\begin{contribution}

All authors contributed equally.


\end{contribution}

\bibliography{sample701}{}
\bibliographystyle{aasjournalv7}



\end{document}